\documentclass[twocolumn,showpacs,preprintnumbers,amsmath,amssymb]{revtex4}

\usepackage{graphicx}
\usepackage{dcolumn}
\usepackage{bm}

\begin{document}

\title{Initial correlation in a system of a spin coupled to a spin bath \\through an intermediate spin }

\author{V.~Semin}
 \email{semin@ukzn.ac.za}
\affiliation{Quantum Research Group, School of Chemistry and Physics,
 University of KwaZulu-Natal, Durban, 4001, South Africa}

\author{I.~Sinayskiy}
 \email{sinayskiy@ukzn.ac.za}
\affiliation{
Quantum Research Group, School of Chemistry and Physics and National Institute
for Theoretical Physics,
 University of KwaZulu-Natal, Durban, 4001, South Africa}

\author{F.~Petruccione}
 \email{petruccione@ukzn.ac.za}
\affiliation{
Quantum Research Group, School of Chemistry and Physics and National Institute
for Theoretical Physics,
 University of KwaZulu-Natal, Durban, 4001, South Africa}

\date{\today}

\begin{abstract}
The strong system-bath correlation is a typical initial condition in many condensed matter and some quantum optical systems. Here, the dynamics of a   spin interacting with a spin bath through an intermediate spin are studied. Initial correlations between the   spin and the intermediate spin are taken into account. The exact analytical expression for the evolution operator of the   spin is found. Furthermore, correlated projection
superoperator techniques are applied to the model and a time-convolutionless master equation to second order is derived. 
 It is shown that the time-convolutionless master equation to second order reproduces the exact dynamics for time-scales of the order $1/\gamma,$ where  $\gamma$ is the coupling of the central spin to the intermediate spin. It is found that there is a strong dependence on the initial system-bath correlations in the dynamics of the reduced system, which cannot be neglected. 
\end{abstract}

\pacs{03.65.-w, 42.50.Ar, 03.65.Yz, 75.10.Jm}
\maketitle

\section{INTRODUCTION}
 The   open quantum dynamics  \cite{toqs} of spin systems finds wide application in various fields of physics, e.g., quantum theory of magnetism \cite{itqss}, quantum information processing \cite{nelsen}, quantum biology \cite{bio} and quantum dots \cite{QD1,QD2,QD3}. Several models were proposed to study decoherence of single and multi-spin systems
interacting with a surrounding environment \cite{1}. The model considered here is quite typical  for various situations, e.g., coupled quantum dots with one of them strongly interacting with an external spin environment or a system of two spin-spin interacting electrons with one of them strongly coupled to surrounding nuclear spins.  Very often, the derivation of the reduced
dynamics involves complications and difficulties that can be overcome in many cases by the
application of approximation techniques. In particular, the Markovian approximation together with the quantum master equation approach turns out to be very useful \cite{2,3}.
However, any approximation method is inevitably based on some assumptions which do
not necessarily reflect the actual properties of the composite system. Moreover, many
realistic spin systems exhibit non-Markovian behaviour for which the standard derivation of the quantum master equation ceases to be applicable. The non-Markovian dynamics of
a   spin-system coupled to a spin environment has been extensively investigated \cite{4, 5, 6, 7,8, oldTCL}.

There are only few solvable models of open quantum system dynamics, which allow for the exact analytical dynamics of the reduced system. Some examples are the damped harmonic oscillator \cite{HPZ}, the spontaneous emission of a two level atom into a zero temperature bath \cite{GAR}, the pure decoherence of a two-level system \cite{UNRUH, PSE} and spin star models  \cite{oldTCL}.  From this point of view, developing new exactly  solvable non-perturbative models plays a crucial role for the deeper understanding of the realistic systems dynamics. Furthermore, exactly solvable models make it possible to test and develop new approximation techniques. A typical spin environment is characterized by  non-Gaussian fluctuations and strong memory effects for a wide range of parameters \cite{stamp}.  Even for the simplest case of a two-level system interacting with a bath of spins in a spin star configuration  the Markov limit for the quantum master equation does not exists  \cite{oldTCL, NM}. Recently, a correlated projection operator approach was  developed \cite{CRL, CRL0, CRL2}. This approximation technique was shown to be an efficient tool in the description of various physical systems \cite{CRL2, CRL3, CRL4}.

Typically, in the derivation of the quantum master equation it is assumed that initially the system and the bath are uncorrelated. However, this is not the case in many condensed matter and quantum optical systems. Taking into account correlations between system and bath gives raise to inhomogeneous terms in the quantum master equation \cite{toqs}. Recently, initial correlations in the Jaynes-Cummings model were studied   using the trace distance \cite{IC1}. It was found that initial correlation plays an important role in the reduced dynamics of the two-level system. The influence of initial correlation on the dissipation of a qubit interacting with a bosonic reservoir at zero temperature was studied in Ref. \cite{IC2}. The role of initial correlations in the decoherence process of a qubit interacting with a bosonic bath for different bath spectral densities was investigated in  \cite{IC3,IC4}, and it was shown that the evolution of the diagonal elements of the reduced system exhibits a  strong dependence on initial correlations. The exact non-Markovian dynamics of open quantum systems in the presence of initial system-reservoir correlations for a photonic cavity system coupled to a general non-Markovian reservoir \cite{IC5} shows that the initial two-photon correlation between the cavity and the reservoir can induce nontrivial squeezing dynamics to the cavity field. However, in all present research the role of initial system bath correlation was investigated for  previously known exactly solvable models \cite{toqs}.

In this paper, we study the reduced dynamics of a
spin coupled to a spin bath through an intermediate spin. The main goals of this work are the following:
Firstly, to use the exact solution of the model derived here to show the importance of initial system-bath correlations. Secondly, to demonstrate that in the case where initial system-bath correlations are present, a rigorously derived master equation will contain an inhomogeneous term which substantially affects the dynamics of the reduced system. Thirdly, by comparing the exact and approximate solutions of the master equation, to understand the limitations of the approximate correlated projector method in the presence of an inhomogeneous term in the quantum master equation.

This paper is organized as follows. In Sec. II we describe the
model of a   spin coupled to a spin bath through an intermediate spin. In Sec. III we present the analytical solution for
the evolution operator and build the reduced density matrix for the   spin. In Sec IV we apply the correlation projection operator method to obtain the quantum master equation in the time-convolutionless form (TCL2) and compare approximate and exact solution of the studied model. Finally, in Sec.
V we discuss the results and conclude.

\section{MODEL}

\begin{figure}
\includegraphics[scale=0.3]{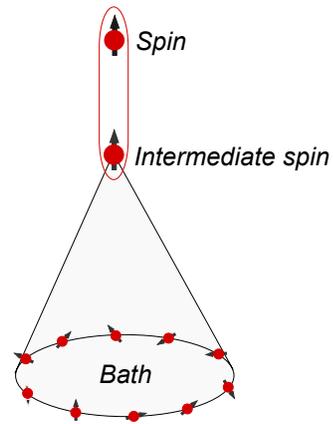}
\caption{(Color online) The model of a spin coupled to a spin bath through 
an intermediate spin. }\label{mod}
\end{figure}

We consider the model of a spin coupled to a spin bath through 
an intermediate spin (see Fig.~\ref{mod}). The interaction Hamiltonian is given by \begin{equation}\label{ham}
H=H_{SI}+H_{IB},\end{equation}
where $$H_{SI}=\frac{\gamma}{2}\left(\sigma_+\tau_-+\sigma_-\tau_+\right)$$
 is the Hamiltonian describing spin-spin interactions and $\sigma_{\pm}, \tau_{\pm}$ 
are the creation (annihilation) operators for the   spin and the intermediate one, respectively. The parameter  $\gamma$ denotes the strength of the spin-spin interaction. The second term in the interaction Hamiltonian describes the interactions between the intermediate spin and the spins in the bath 
\begin{equation}
H_{IB}=\frac{\alpha}{2\sqrt{N}}\left(\tau_+J_-+\tau_-J_+\right).\nonumber\end{equation}
In the above equation, $J_{\pm}=\sum\limits_{i=1}^{N}\sigma^i_\pm$, and  $\sigma^i_\pm$  are creation (annihilation) operators of $i$-th spin of the bath, $\alpha$ is the strength of the interaction and $N$ denotes the numbers of bath spins. The factor $1/\sqrt{N}$ is introduced in the above Hamiltonian as usual \cite{yamen} to obtain the correct behaviour in the thermodynamic limit ($N\rightarrow\infty$). The uniform coupling of the Hamiltonian $H_{IB}$ is a simplification allowing to build an analytical solution for the model. In this paper units are chosen such that
$k_B=\hbar=1$. 

\section{EXACT SOLUTION}

To describe the exact dynamics of the total system we need to specify an initial state of the total system which is given by the  density operator $\rho_{\mathrm{tot}}(0)$ and to find an evolution operator of the total system $U(t)$ in an explicit form, as
\begin{equation}U(t)=\exp[-i H t]\label{U}.\end{equation} 
With the knowledge of the evolution operator and the initial state of the total system the reduced dynamics of the   spin can be found as  
\begin{equation}\label{reduced}
\rho^{S}(t)=\mathrm{tr}_{IB}\{U(t)\rho_{\mathrm{tot}}(0)U^{\dagger}(t)\}.
\end{equation} 
Here we will consider an initially correlated state between the   spin and the intermediate spin, while the rest of the bath is assumed to be unpolarized.

The initial state of the total system reads,
\begin{equation}\label{initial}
\rho_{\mathrm{tot}}(0)=\rho_{SI}(0)\otimes \rho_{B}(0),
\end{equation} 
where $\rho_{SI}(0)$ is the density matrix of the spin and the intermediate one is given by a generic $X$-like initial two-qubit density matrix, as
\begin{equation}\label{initial2}
\rho_{SI}(0)=\left( 
\begin{array}{cccc}
\rho_{11}^0 & 0 & 0 & \rho_{14}^0 \\
0&\rho_{22}^0 & \rho_{23}^0 & 0 \\
0& \rho_{32}^0 & \rho_{33}^0 & 0 \\
\rho_{41}^0 & 0 & 0 & \rho_{44}^0 \\
\end{array}
\right),
\end{equation} 
while  $\rho_B(0)$ is the density matrix describing the unpolarized state of $N$ particle with spin 1/2, as

\begin{equation}
\rho_{B}(0)=\frac{I_B}{2^N}.
\end{equation} 
The density matrix \eqref{initial2} contains any two spin states of the form
\begin{eqnarray}
|\Phi^{\pm}_{\theta,\beta}\rangle=\sin(\theta)|+\rangle_S|-\rangle_I\pm\exp(i\beta)\cos(\theta)|-\rangle_S|+\rangle_I,\label{phi}\\ |\Psi^{\pm}_{\theta,\beta}\rangle=\sin(\theta)|+\rangle_S|+\rangle_I\pm\exp(i\beta)\cos(\theta)|-\rangle_S|-\rangle_I, \label{psi}
\end{eqnarray}
where the $|\pm\rangle_S$ and
$|\pm\rangle_I$ are the eigenvectors of $\sigma_z$ and $\tau_z$, respectively. We notice that $|\Phi^{\pm}_{\frac{\pi}{4},0}\rangle$ and  $|\Psi^{\pm}_{\frac{\pi}{4},0}\rangle$
are the usual Bell states.

\subsection{Analytical expression for the evolution operator}
Let $U_{ij}$ denote the components of the evolution operator $U$ in the basis $\{|\!--\rangle,\,\,|\!-+\rangle,\,\,|\!+-\rangle,\,\,|\!++\rangle\}$ of eigenvectors of the operator $\sigma_z\otimes\tau_z,$ where $\sigma_z$ is a diagonal operator of the   spin, and $\tau_z$ is a diagonal operator of the intermediate spin. We can write
\begin{eqnarray}
U|\!\!+\!\!+\rangle=U_{11}|\!\!+\!\!+\rangle+U_{21}|\!\!+\!\!-\rangle+U_{31}|\!\!-\!\!+\rangle+U_{41}|\!\!-\!\!-\rangle, \label{basicU1}\\
U|\!\!+\!\!-\rangle=U_{12}|\!\!+\!\!+\rangle+U_{22}|\!\!+\!\!-\rangle+U_{32}|\!\!-\!\!+\rangle+U_{42}|\!\!-\!\!-\rangle,\label{basicU2}\\
U|\!\!-\!\!+\rangle=U_{13}|\!\!+\!\!+\rangle+U_{23}|\!\!+\!\!-\rangle+U_{33}|\!\!-\!\!+\rangle+U_{43}|\!\!-\!\!-\rangle,\label{basicU3}\\
U|\!\!-\!\!-\rangle=U_{14}|\!\!+\!\!+\rangle+U_{24}|\!\!+\!\!-\rangle+U_{34}|\!\!-\!\!+\rangle+U_{44}|\!\!-\!\!-\rangle\label{basicU4}.
\end{eqnarray}
On the other hand, the operator $U$ satisfies the Schr\"{o}dinger equation
\begin{equation} \label{shred}
i \frac{d}{dt}U|i,j\rangle=HU|i,j\rangle,\\
\end{equation}  
where $i,j = + \,\, \text{or}\,\,-$.

Substituting Eqs. (\ref{basicU1})-(\ref{basicU4}) into Eq. (\ref{shred}) yields the following system of coupled differential equations
\begin{equation}
\left\lbrace 
\begin{array}{lll}\label{system_U}
i \dot{U}_{1j}=\frac{\alpha}{2\sqrt{N}}J_-U_{2j}, \\
i \dot{U}_{2j}=\frac{\alpha}{2\sqrt{N}}J_+U_{1j}+\frac{\gamma}{2}U_{3j}, \\
i \dot{U}_{3j}=\frac{\alpha}{2\sqrt{N}}J_-U_{4j}+\frac{\gamma}{2}U_{2j}, \\
i \dot{U}_{4j}=\frac{\alpha}{2\sqrt{N}}J_+U_{3j}. \\
\end{array}
\right.
\end{equation} 
Here, $j=1,\, 2,\, 3,\, 4$ is the number of the column of the evolution operator $U$ in the chosen basis.

Differentiating the second and third equations in  \eqref{system_U} and combining with the first and fourth equations in  \eqref{system_U} we obtain
\begin{equation}
\left\lbrace 
\begin{array}{lll} \label{bassyst}
i \dot{U}_{1j}=\frac{\alpha}{2\sqrt{N}}J_-U_{2j}, \\
i \ddot{U}_{2j}=-i\frac{\alpha^2}{4N}J_+J_-U_{2j}+\frac{\gamma}{2}\dot{U}_{3j}, \\
i \ddot{U}_{3j}=-i\frac{\alpha^2}{4N}J_-J_+U_{3j}+\frac{\gamma}{2}\dot{U}_{2j}, \\
i \dot{U}_{4j}=\frac{\alpha}{2\sqrt{N}}J_+U_{3j}. \\
\end{array}
\right. 
\end{equation} 
All terms in this system of the differential equations are diagonal in the common eigenbasis  of the $J^2$ and $J_z$ operators of the bath. Hence, the standard method of solving systems of differential equations can be applied.

Initial conditions for the evolution operator follow from its definition \eqref{U}. Clearly, we have
\begin{equation} 
\frac{\partial^n}{\partial t^n} U(t)\mid_{t\rightarrow 0} =(-iH)^n.
\end{equation} 
Using the previous relation we can determine that the system \eqref{bassyst} admits the following solution 
\begin{widetext}
\begin{equation}\label{exactsolut}
\begin{array}{lll}
U_{11}=1-\frac{\alpha^2}{4 N}J_-\left\lbrace \left( -1+\cosh\left(A_-\right)\right) C_+ F^{-1}G_-^{-2}-\left( -1+\cosh\left(A_+\right)\right) C_- F^{-1}G_+^{-2}\right\rbrace J_+ , \\
U_{12}=\frac{i \alpha}{2\sqrt{2N}}J_-F^{-1}\left\lbrace \sinh(A_+)C_-G_+^{-1}- \sinh(A_-)C_+G_-^{-1}\right\rbrace, \\
U_{13}=\frac{\alpha \gamma}{4\sqrt{N}}J_-F^{-1}\left\lbrace-\cosh(A_+)+\cosh(A_-) \right\rbrace,\\ 
U_{14}=\frac{i\sqrt{2}\alpha^2\gamma}{8N}J_-F^{-1}\left\lbrace \sinh(A_+)G_+^{-1}-\sinh(A_-)G_-^{-1}\right\rbrace J_-, \\
U_{21}=\frac{i\alpha}{2\sqrt{2N}}F^{-1}\left\lbrace \sinh\left(A_+\right) C_-G_+^{-1}-\sinh\left(A_-\right) C_+ G_-^{-1}\right\rbrace J_+ , \\
U_{22}=\frac{1}{2}F^{-1}\left\lbrace- \cosh(A_+)C_-+ \cosh(A_-)C_+\right\rbrace, \\
U_{23}=\frac{i\gamma}{2\sqrt{2}}F^{-1}\left\lbrace-\sinh(A_+)G_++\sinh(A_-)G_- \right\rbrace,\\ 
U_{24}=\frac{\alpha\gamma}{4\sqrt{N}}F^{-1}\left\lbrace -\cosh(A_+)+\cosh(A_-)\right\rbrace J_-,\\
U_{31}=\frac{\alpha\gamma}{4\sqrt{N}}F^{-1}\left\lbrace -\cosh(A_+)+\cosh(A_-)\right\rbrace J_+,\\
U_{32}=\frac{i\gamma}{2\sqrt{2}}F^{-1}\left\lbrace-\sinh(A_+)G_++\sinh(A_-)G_- \right\rbrace,\\ 
U_{33}=\frac{1}{2}F^{-1}\left\lbrace- \cosh(A_+)C_-+ \cosh(A_-)C_+\right\rbrace, \\
U_{34}=\frac{i\alpha}{2\sqrt{2N}}F^{-1}\left\lbrace \sinh\left(A_+\right) C_-G_+^{-1}-\sinh\left(A_-\right) C_+ G_-^{-1}\right\rbrace J_- , \\
U_{41}=\frac{i\sqrt{2}\alpha^2\gamma}{8N}J_+F^{-1}\left\lbrace \sinh(A_+)G_+^{-1}-\sinh(A_-)G_-^{-1}\right\rbrace J_+, \\
U_{42}=\frac{\alpha \gamma}{4\sqrt{N}}J_+F^{-1}\left\lbrace-\cosh(A_+)+\cosh(A_-) \right\rbrace,\\ 
U_{43}=\frac{i \alpha}{2\sqrt{2N}}J_+F^{-1}\left\lbrace \sinh(A_+)C_-G_+^{-1}- \sinh(A_-)C_+G_-^{-1}\right\rbrace, \\
U_{44}=1-\frac{\alpha^2}{4 N}J_+\left\lbrace \left( -1+\cosh\left(A_-\right)\right) C_+ F^{-1}G_-^{-2}-\left( -1+\cosh\left(A_+\right)\right) C_- F^{-1}G_+^{-2}\right\rbrace J_- . \\
\end{array}
\end{equation} 
\end{widetext}
In the above system of equations we have used the notation
\begin{equation}
\begin{array}{lll}\label{GF}
G_\pm=1/2\sqrt{-\frac{\alpha^2}{N}(J_+J_-+J_-J_+)-\gamma^2\pm 4F},\\
F=\frac{1}{4}\sqrt{4\frac{\alpha^4}{N^2}J_z^2+2\frac{\alpha^2\gamma^2}{N}(J_+J_-+J_-J_+)+\gamma^4},\\
C_\pm=(2\alpha^2J_z\pm 4F N+\gamma^2 N)/(4N), \\
A_\pm=G_\pm t/\sqrt{2}.
\end{array}
\end{equation} 

\subsection{Reduced density matrix}

Using the exact analytical expression for the evolution operator \eqref{exactsolut} after the trace over the intermediate spin variables we find explicitly the reduced dynamics of the spin (see Eq. \eqref{reduced}), as

\begin{eqnarray}
\rho^S_{11}(t)=&\!\!\!\!\mathrm{Tr}_B\left[ \rho^0_{11}U_{11}U_{11}^\dagger+(\rho^0_{22}U_{12}+\rho^0_{32}U_{13})U_{12}^\dagger\right. \nonumber\\
+&\!\!\!\!\!\!\!\!\!\!\!\!\!\!\!\!\!\!\!\rho^0_{44}U_{14}U_{14}^\dagger+(\rho^0_{23}U_{12}+\rho^0_{33}U_{13})U_{13}^\dagger \nonumber\\
+&\!\!\!\!\!\!\!\!\!\!\!\!\!\!\!\!\!\!\!\rho^0_{11}U_{21}U_{21}^\dagger+(\rho^0_{22}U_{22}+\rho^0_{32}U_{23})U_{22}^\dagger\label{r11}\\
+&\!\!\!\left.\rho^0_{44}U_{24}U_{24}^\dagger+(\rho^0_{23}U_{22}+\rho^0_{33}U_{23})U_{23}^\dagger\right]/2^N\nonumber, \\
\rho^S_{22}(t)=&\!\!\!\!\!\!\!\!\!\!\!\!\!\!\!\!\!\!\!\!\!\!\!\!\!\!\!\!\!\!\!\!\!\!\!\!\!\!\!\!\!\!\!\!\!\!\!\!\!\!\!\!\!\!\!\!\!\!\!\!\!\!\!\!\!\!\!\!\!\!\!\!\!\!\!\!\!\!1-\rho^S_{11}(t),\nonumber \\
\rho^S_{12}(t)=&\!\!\!\!\!\!\!\!\!\!\!\!\!\!\!\!\!\!\!\!\!\!\!\!\!\!\!\!\!\!\!\!\!\!\!\!\!\!\!\!\!\!\!\!\!\!\!\!\!\!\!\!\!\!\!\!\!\!\!\!\!\!\!\!\!\!\!\!\!\!\!\!\!\!\!\!\!\rho^S_{21}(t)=0.\label{r12}
\end{eqnarray}
The relation \eqref{exactsolut}, \eqref{r11}, \eqref{r12} are the main result of this article.

To perform the trace  over the collective bath variables one needs to take into account the degeneracy of   states of the bath as
\begin{equation}\label{tr}
\mathrm{tr}_B\rho=\sum\limits_{j=0,\frac{1}{2}}^{N/2}\sum\limits_{m=-j}^{j}\nu(N,j) \langle j,m|\rho|j,m\rangle,
\end{equation}
where the degeneracy is given by
\begin{equation} \label{Def-N-J1}
\nu(N,j)=
 \binom{N}{\frac{N}{2}+j}-\binom{N}{\frac{N}{2}+j+1}.
\end{equation} 
The vectors $|j,m\rangle$ are eigenvectors of the bath operators $J^2$ and $J_z$, namely
\begin{eqnarray}\label{act}
J_z|j,m\rangle=m|j,m\rangle,\\
J^2|j,m\rangle=j(j+1)|j,m\rangle,\\
J_\pm|j,m\rangle=\sqrt{j(j+1)-m(m\pm 1)}|j,m \pm 1\rangle,
\end{eqnarray} 
where the eigenvalues vary from $j=1/2$ to $N/2$ for odd $N,$ or from $j=0$ to $N/2$ for even $N$, and $m=-j,...,j$.

\begin{figure}[t]
\includegraphics[scale=0.6]{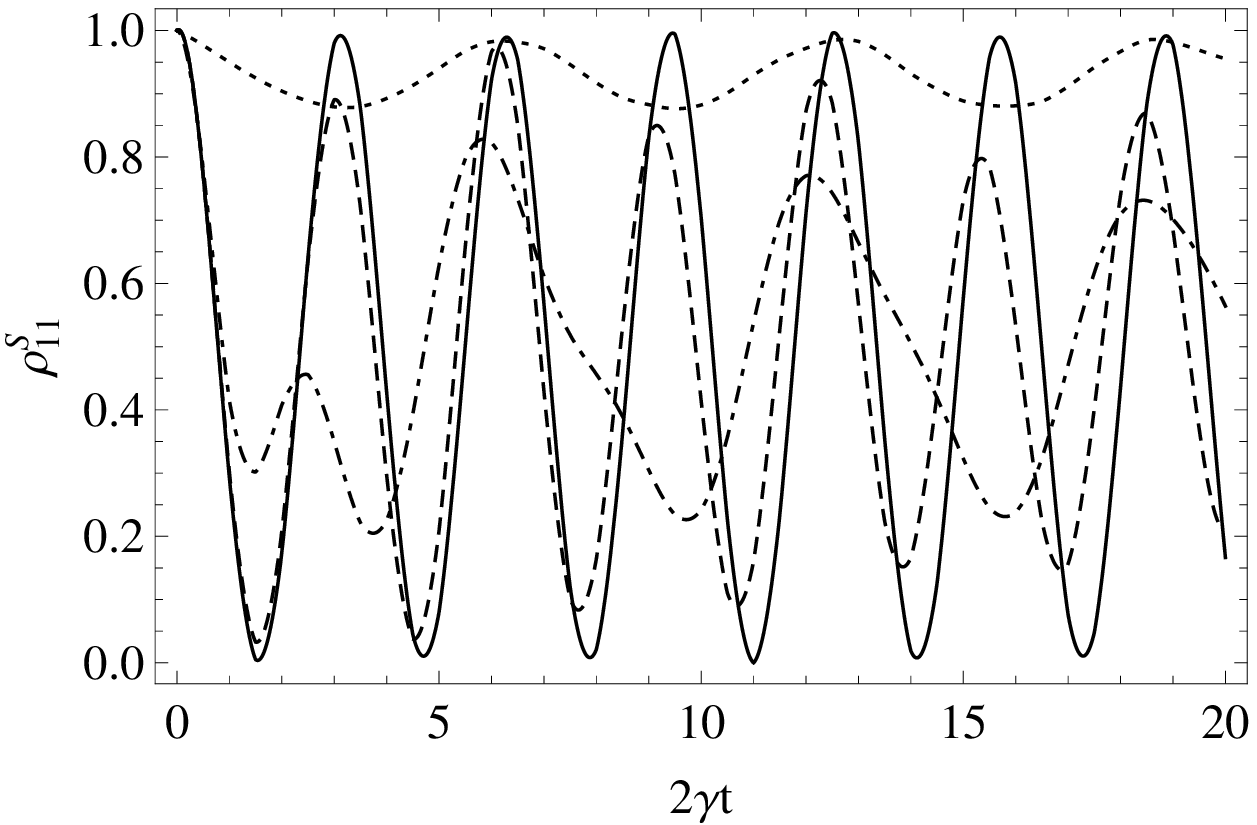}
\caption{Dynamics of the population of the upper state $\rho^S_{11}$ based on the exact solution of the Schr\"{o}dinger equation  \eqref{r11} for different coupling strengths between intermediate spin and environmental spins. Solid, dashed, dot-dashed and dotted lines correspond to $\alpha=0$, $\alpha=\gamma/4$, $\alpha=\gamma$  and  $\alpha=100\gamma$, respectively. The initial state is the Bell state $|\Phi^+_{\pi/2,0}\rangle$; number of spins in the bath $N=50$.}\label{Exact_Cor1}
\end{figure}

\begin{figure}
\includegraphics[scale=0.6]{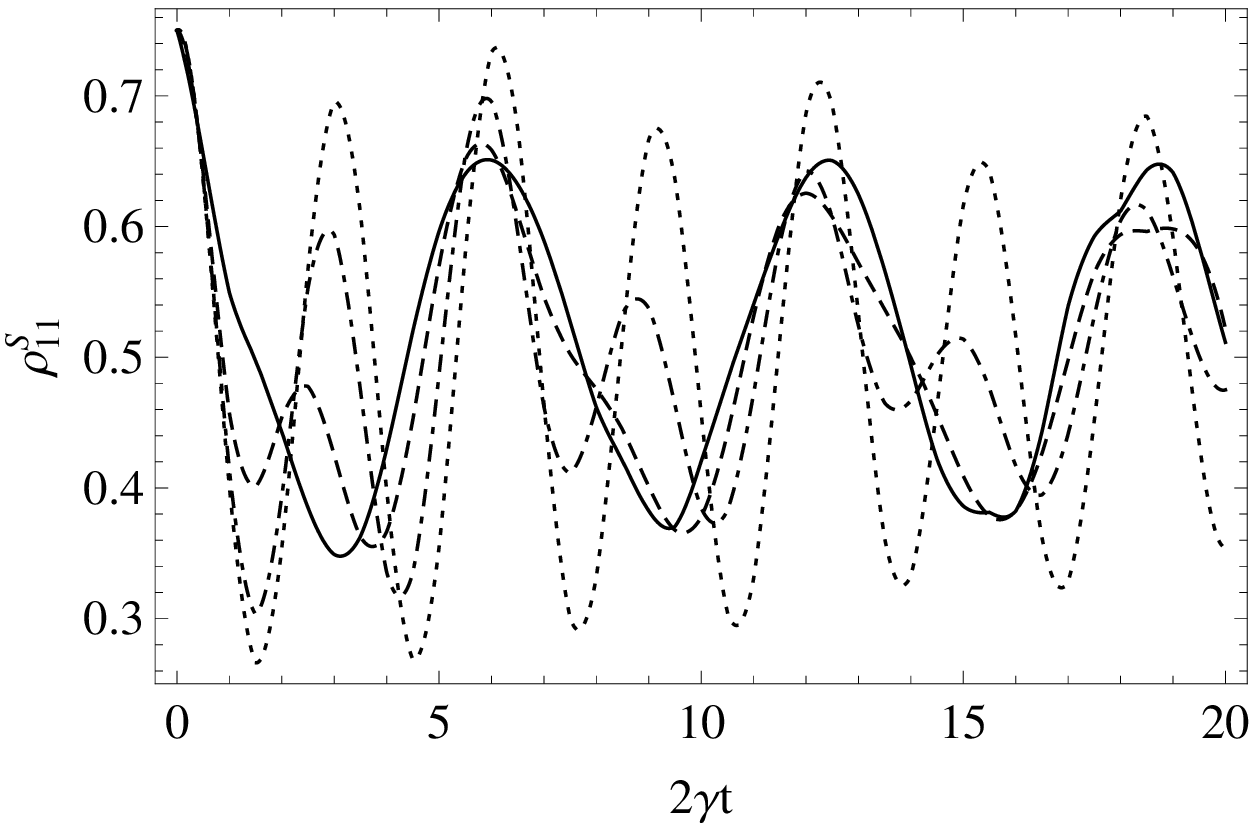}
\caption {Dynamics of the population of the upper state $\rho^S_{11}$ based on the exact solution of the Schr\"{o}dinger equation  \eqref{r11} for different coupling strengths between intermediate spin and environmental spins. Solid, dashed, dot-dashed and dotted lines correspond to $\alpha=2\gamma$, $\alpha=\gamma$, $\alpha=\gamma/2$  and  $\alpha=\gamma/4$, respectively.  The initial state is the Bell state $|\Phi^+_{\pi/3,0}\rangle$;  number of  spins in the bath $N=50$.}\label{Exact_Cor}
\end{figure}

Using the exact expression for the reduced density matrix, we can analyse the  dynamics of the   spin. The dynamics  of the probability to find the spin in the state $|+\rangle$ is shown in Figs.~ \ref{Exact_Cor1} -- \ref{Exact_len}.

\subsection{The limit of large number of bath spins ($N\rightarrow \infty$)}

This subsection is devoted to the case of an infinite number of spins in the bath, i.e., the case $N\rightarrow \infty$.   The limit of large number of spins can be obtained using the following formula \cite{yamen}
\begin{eqnarray}
&\lim\limits_{N\rightarrow \infty}2^{-N}\mathrm{tr}_B\left\lbrace f\left(\frac{J_\pm J_\mp}{N},\frac{J_z}{\sqrt{N}} \right)  \right\rbrace \label{lim} \\
=& \left( \frac{2}{\pi}\right) ^{3/2}\int\limits_{-\infty}^{\infty}dm \int\limits_{\mathbf{C}}dzdz^*f(|z|^2,m)e^{-2(m^2+|z|^2)}.\nonumber
\end{eqnarray}

An exact calculation shows that
\begin{widetext}
\begin{eqnarray}
\rho_{11}^{S}(t)_{N\rightarrow \infty}=1/2\left(1+(\rho_{11}^0+\rho_{22}^0-\rho_{33}^0-\rho_{44}^0)\cos\left( \frac{\gamma t}{2}\right) +i (\rho_{23}^0-\rho_{32}^0)\sin\left( \frac{\gamma t}{2}\right) \right)& \nonumber  \\
-\left( (\rho_{11}^0+\rho_{22}^0-\rho_{33}^0-\rho_{44}^0)\cos\left( \frac{\gamma t}{2}\right)+i (\rho_{23}^0-\rho_{32}^0)\sin\left( \frac{\gamma t}{2}\right)\right)f(t) \label{limex}\\
+\left( (\rho_{11}^0-\rho_{22}^0+\rho_{33}^0-\rho_{44}^0)\sin\left( \frac{\gamma t}{2}\right)+i (\rho_{23}^0-\rho_{32}^0)\cos\left( \frac{\gamma t}{2}\right)\right)g(t),\nonumber
\end{eqnarray}
\begin{equation}
\rho_{22}^{S}(t)_{N\rightarrow \infty}=1-\rho_{11}^{S}(t)_{N\rightarrow \infty},
\end{equation}
\end{widetext}
where the functions $f(t)$ and $g(t)$ are given by

\begin{equation}
f(t)=\frac{\gamma^2}{\alpha^2}e^{\gamma^2/2\alpha^2}\sum _{n=1}^{\infty } \frac{(\gamma  t)^{2 n}}{2^{n+2}(2 n)!} \frac{\partial ^{n-1}}{\partial A^{n-1}}\frac{e^{-A/2}}{A}\lvert_{A\rightarrow\gamma^2/\alpha^2},\nonumber
\end{equation}
and
\begin{eqnarray}
g(t)=&\!\!\!\!\!\!\!\!\!\!\!\!\!\!\!\!\!\!\!\!\!\!\!\!\!\!\!\!\!\!\!\!\!\!\!\!\frac{i \gamma}{4\alpha}e^{\gamma^2/2\alpha^2- \alpha^2 t^2/8}\sqrt{\frac{\pi }{2}}\nonumber\\ 
&\times \left( \mathrm{erf}\left(\frac{2\gamma-i  \alpha^2 t}{2\sqrt{2}  \alpha}\right)-\mathrm{erf}\left(\frac{2\gamma+i \alpha^2 t}{2\sqrt{2}  \alpha}\right)\right).\nonumber
\end{eqnarray}

\begin{figure}
\includegraphics[scale=0.6]{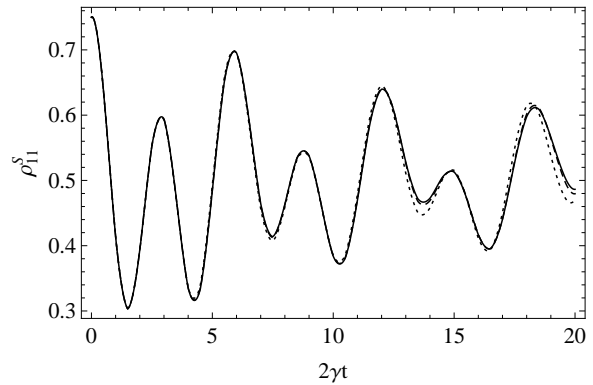}
\caption{Dynamics of the population of the upper state $\rho^S_{11}$ based on the exact solution of the Schr\"{o}dinger equation  \eqref{r11} for different number of spins in the environment.  Dotted, dashed and solid lines correspond to $10$, $50$ and $\infty$ number of the environmental spins, respectively. The initial state is the Bell state $|\Phi^+_{\pi/3,0}\rangle$; coupling strength $\alpha=\gamma/2$.}\label{Exact_N}
\end{figure}

\begin{figure}
\includegraphics[scale=0.6]{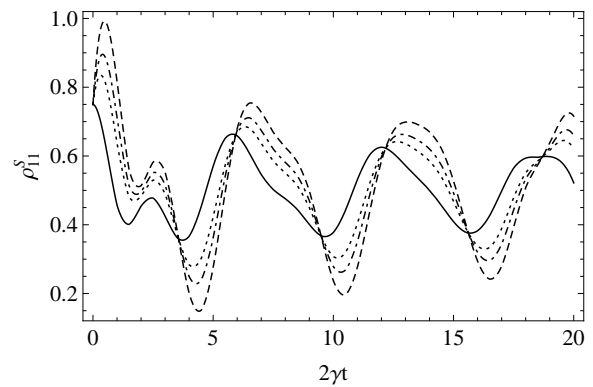}
\caption{Dynamics of the population of the upper state $\rho^S_{11}$ based on the exact solution of the Schr\"{o}dinger equation  \eqref{r11} for different initial system-environment correlations. Solid, dashed, dotted and dot-dashed lines correspond to initial states $|\Phi^+_{\pi/3,0}\rangle$, $|\Phi^+_{\pi/3,\pi/2}\rangle$ , $|\Phi^+_{\pi/3,\pi/4}\rangle$  and  $|\Phi^+_{\pi/3,\pi/6}\rangle$, respectively.  Number of spins in the bath $N=50$; coupling strength $\alpha=\gamma$.}\label{Exact_len}
\end{figure}

\begin{figure}
\includegraphics[scale=0.6]{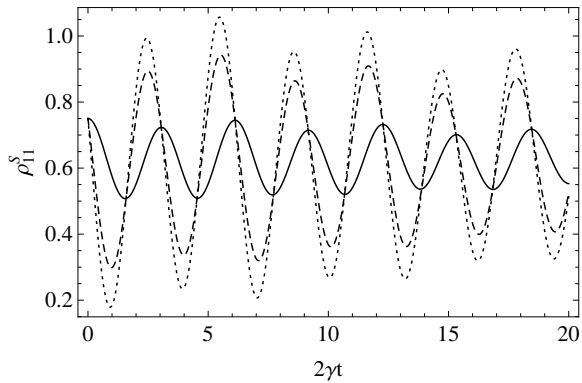}
\caption{Dynamics of the population of the upper state $\rho^S_{11}$ based on the exact solution of the Schr\"{o}dinger equation Eq. \eqref{limex} in the thermodynamic limit for different initial system-environment correlations. Solid, dashed and dotted lines corresponds to initial states $|\Phi^-_{\pi/3,0}\rangle$, $|\Phi^-_{\pi/3,\pi/4}\rangle$  and  $|\Phi^-_{\pi/3,\pi/2}\rangle$,  respectively.  The coupling strength is $\alpha=\gamma/4$.}\label{Nlim}
\end{figure}

Using the exact expression for the reduced density matrix we can analyse the dynamics of the spin in the thermodynamic limit, Eq. \eqref{limex}. The dynamics  of the probability to find the spin in the state $|+\rangle$ is shown in Fig.~ \ref{Exact_N} and Fig. \ref{Nlim}.


\section{APPROXIMATION TECHNIQUES}
\subsection{Projection operator techniques}
Projection operator techniques are a powerful tool of statistical physics \cite{NAKAJIMA,ZWANZIG,Grabert, Kubo, Prigogine}.
The application of projection operator techniques in the theory of open quantum systems is based on  considering  the operation of tracing over the environment as a formal projection $\rho\mapsto\mathcal{P}\rho$ in the state space of the total system \cite{toqs, toqs2}. The superoperator $\mathcal{P}$  has the property of a projection operator, that is $\mathcal{P}^2=\mathcal{P},$ and the density matrix $\mathcal{P}\rho$ is said to be the relevant part of the density $\rho$ of the total system. Correspondingly, 
a projector $\mathcal{I}-\mathcal{P}$ or $\rho\mapsto\mathcal{Q}\rho$ is defined as a projection onto  the irrelevant part of the total density matrix.

\subsection{Time-convolutionless master equation}
One of all possible ways of deriving an exact master equation for the
relevant part of $\rho$ is to remove the dependence of the
system's dynamics on the full history of the system and to
formulate a time-local equation of motion, which is given by
\begin{equation}\label{TCL-GEN}
\frac{\partial}{\partial t}\mathcal{P} \rho(t)=\mathcal{K}(t)\mathcal{P} \rho(t)+\mathcal{I}(t)\mathcal{Q}\rho(0).\end{equation}

This equation is called the time-convolutionless (TCL) master
equation, and $\mathcal{K}(t)$ is a time-dependent superoperator,
which is referred to as the TCL generator. As for the Nakajima-Zwanzig equation \cite{NAKAJIMA,ZWANZIG},
in general there is also an inhomogeneous term proportional to $\mathcal{Q}
\rho(0)$ on the right-hand side of Eq.~(\ref{TCL-GEN}).

The expansion to second order in $H$ of the TCL generator and the inhomogeneity are given by \cite{toqs}
\begin{eqnarray}
\mathcal{K}(t)=\mathcal{P}\mathcal{L}(t)\mathcal{P}+\int\limits_{0}^{t}ds \mathcal{P}\mathcal{L}(t)\mathcal{L}(s)\mathcal{P},\label{generP}\\
\mathcal{I}(t)=\mathcal{P}\mathcal{L}(t)\mathcal{Q}+\int\limits_{0}^{t}ds\mathcal{P}\mathcal{L}(t)\mathcal{L}(s)Q.\label{generQ}\end{eqnarray}

\subsection{Correlated projection superoperators}
The starting point of the projection operator techniques is the introduction
of a superoperator $\mathcal{P}$ which acts on the total system's density
matrix, and which is usually defined by
\begin{equation} \label{STANDARD-PROJECTION}
    \mathcal{P} \rho = (\mathrm{tr}_E\rho) \otimes \rho_0,
\end{equation}
where $\rho_0$ is some fixed state of the environment. Obviously, the map
$\mathcal{P}$ satisfies the condition of a projector, namely
${\mathcal{P}}^2={\mathcal{P}}$.

The complementary map is defined via $\mathcal{Q} = I - \mathcal{P}$,
where $I$ denotes the identity. Note that $\mathcal{P} \rho$
contains all information about the open system in the sense that
for the expectation value of any observable $\mathcal{O}_S$ of the
open system the relation $\mathrm{tr} \{ \mathcal{O}_S \rho \} =
\mathrm{tr} \{ \mathcal{O}_S \mathcal{P} \rho \}$ holds.

The projector \eqref{STANDARD-PROJECTION} is not the only
possible choice \cite{CRL}. In fact, a general class of projection
superoperators can be represented as follows,

\begin{equation} \label{PROJECTION-GENFORM}
 {\mathcal{P}}\rho = \sum_i {\mathrm{tr}}_E \{ A_i \rho \}
 \otimes B_i,
\end{equation}
where $\{A_i\}$ and $\{B_i\}$ are two sets of linear independent
Hermitian operators on ${\mathcal{H}}_E$ satisfying the relations
\begin{eqnarray}
 {\mathrm{tr}}_E \{ B_i A_j \} &=& \delta_{ij}, \label{BjAi} \\
 \sum_i ({\mathrm{tr}}_E B_i) A_i &=& I_E, \label{TRACE-PRESERVING} \\
 \sum_i A_i^T \otimes B_i &\geq& 0. \label{COND-POS}
\end{eqnarray}

Once $\mathcal{P}$ is chosen, the dynamics of the open system is
uniquely determined by the dynamical variables
\begin{equation}
    \rho_i(t) = \mathrm{tr}_E \{ A_i \rho(t) \}.
\end{equation}
The connection to the reduced density matrix is simply given by
\begin{equation}\label{rhosum}
    \rho_S(t) = \sum_i \rho_i(t),
\end{equation}
and the normalization condition reads
\begin{equation}
    \mathrm{tr}_S \> \rho_S(t) = \sum_i \mathrm{tr}_S \> \rho_i(t) = 1.
\end{equation}

The TCL equation (\ref{TCL-GEN}) together with the projection operator \eqref{PROJECTION-GENFORM} leads to a
coupled system of time-local differential equations,
\begin{equation} \label{TCL-RHO-I}
 \frac{d}{dt} \rho_i(t) = \sum_j \mathcal{K}_{ij}(t) \rho_j(t)+\sum_j \mathcal{I}_{ij}(t) \rho_j(0),
\end{equation}
with superoperators defined by
\begin{eqnarray}
\mathcal{K}_{ij}& \!\!\!(t) \mathcal{O}_S \equiv
 {\mathrm{tr}}_E \left\{ A_i \mathcal{K}(t) (\mathcal{O}_S \otimes B_j) \right\},\\
\mathcal{I}_{ij}&\!\!\!(t) \mathcal{O}_S \equiv
  {\mathrm{tr}}_E \left\{ A_i \mathcal{I}(t) (\mathcal{O}_S \otimes B_j) \right\}.
\end{eqnarray}
In the subsequent discussion we assume that $\mathcal{K}$ and $\mathcal{I}$ are defined through Eqs. \eqref{generP} and \eqref{generQ}

\subsection{Application to the model}
As it was indicated by Fisher and Breuer \cite{CRL} the correlated projection approach is most efficient if  projections on subspaces corresponding to some conserving quantity are considered. For the spin-star models the most appropriate conserving quantity is the $z$-projection of the total angular momentum of the total system. Explaining the symmetries of the model under consideration we generalise the correlated projection operator as follows

\begin{eqnarray} \label{Def-N-J}
\mathcal{P}\rho&=&\sum\limits_{j}\sum\limits_{m=-j}^{j}\mathrm{tr}_{IB}\left(\Pi^+_{jm}\rho\right)\otimes \frac{\Pi^+_{jm}}{N_j}  \\ \label{ourcor}&&+\sum\limits_{j}\sum\limits_{m=-j}^{j}\mathrm{tr}_{IB}\left( \Pi^-_{jm}\rho\right) \otimes \frac{\Pi^-_{jm}}{N_j}, \nonumber
\end{eqnarray}
where $\Pi^\pm_{jm}=|\pm\rangle\langle \pm|\otimes |j,m\rangle\langle j,m|.$  The $|\pm\rangle$ are eigenvectors $\tau_z$ and $|j,m\rangle$ are eigenvectors of the bath operators  $J_z$ and $J^2$. The corresponding eigenvalues are $m$ and $j(j+1)$. We introduce the notation

\begin{equation}
 \qquad N_j =\mathrm{tr}\Pi^\pm_{jm}=
 \binom{N}{\frac{N}{2}+j}-\binom{N}{\frac{N}{2}+j+1}.
\end{equation}
Naturally, the projection on the irrelevant part is defined as

\begin{equation}\label{irrelev}
\mathcal{Q}\rho=\rho-\mathcal{P}\rho.
\end{equation}

Combining Eqs.~\eqref{Def-N-J},\eqref{irrelev} and \eqref{TCL-RHO-I} with the Hamiltonian \eqref{ham} and for the initial state \eqref{initial}, using (30) and (31), we get a system of differential equations for the quantities  $R_{jm}=\mathrm{tr}_{IB}\left(\Pi^+_{jm}\rho\right),$ and $r_{jm}=\mathrm{tr}_{IB}\left(\Pi^-_{jm}\rho\right),$ namely
\begin{eqnarray}
\dot{r}_{jm}&&\!\!\!\!\!\!=-\frac{\alpha^2}{2N}b(j,-m)(r_{jm}-R_{jm-1})t \label{1} \\ &&-\frac{\gamma^2}{4}(r_{jm}\sigma^+\sigma^-+\sigma^+\sigma^-r_{jm}
-2\sigma^+R_{jm}\sigma^-)t+\Lambda_1,\nonumber \\
\dot{R}_{jm}&&\!\!\!\!\!\!=-\frac{\alpha^2}{2N}b(j,m)(R_{jm}-r_{jm+1})t \label{2} \\&&-\frac{\gamma^2}{4}(R_{jm}\sigma^-\sigma^++\sigma^-\sigma^+R_{jm}
-2\sigma^-r_{jm}\sigma^+)t+\Lambda_2,\nonumber
\end{eqnarray}
where $b(j,m)=(j-m)(j+m+1).$ 
The inhomogeneities in the above equation are defined as 
\begin{equation}
\Lambda_1=\left( 
\begin{array}{ccc}
i\frac{\gamma}{2^{N+1}}(\rho_{23}^0-\rho_{32}^0)&&0\\
0&&0
\end{array}
\right) ,
\end{equation} 

\begin{equation}
\Lambda_2=\left( 
\begin{array}{ccc}
0&0 \\
0&-i\frac{\gamma}{2^{N+1}}(\rho_{23}^0-\rho_{32}^0)\\
\end{array}
\right).
\end{equation} 
The initial conditions for Eqs.~\eqref{1} and ~\eqref{2} are given by

\begin{equation}
R_{jm}(0)=N_j/2^N\left( 
\begin{array}{ccc}
\rho_{11}^0&&0\\
0&&\rho_{33}^0
\end{array}
\right) ,
\end{equation} 

\begin{equation}
r_{jm}(0)=N_j/2^N\left( 
\begin{array}{ccc}
\rho_{22}^0&&0\\
0&&\rho_{44}^0
\end{array}
\right).
\end{equation} 

From Eqs.~ \eqref{1} and \eqref{2} we find a closed system of equations for the matrix elements
\begin{eqnarray}
\dot{R}^{11}_{jm-1}=&\!\!\!\!\!\!\!\!\!\!\!\!\!\!\!\!\!\!\!\!\!\!\!\!\frac{\alpha^2}{2N}b(j,-m)(r^{11}_{jm}-R^{11}_{jm-1})t, \label{tl1} \\ \dot{r}^{11}_{jm}=&\!\!\!\!\!\!\!\!\!\!\!\!\!\!\!\!\!\!\!\!\!\!\!\!\!\!\!-\frac{\alpha^2}{2N}b(j,-m)(r^{11}_{jm}-R^{11}_{jm-1})t \label{tl2} \\&+\frac{\gamma^2}{2}(R_{jm}^{22}-r_{jm}^{11})t+i\frac{\gamma}{2^{N+1}}(\rho_{23}^0-\rho_{32}^0),\nonumber \\
\dot{R}^{22}_{jm}=&\!\!\!\!\!\!\!\!\!\!\!\!\!\!\!\!\!\!\!\!\!\!\!\!\!\!\!\!\!\!\!\!2\frac{\alpha^2}{N}b(j,m)(r^{22}_{jm+1}-R^{22}_{jm})t\label{tl3}\\&-\frac{\gamma^2}{2}(R_{jm}^{22}-r_{jm}^{11})t-i\frac{\gamma}{2^{N+1}}(\rho_{23}^0-\rho_{32}^0),\nonumber \\
\dot{r}^{22}_{jm+1}=&\!\!\!\!\!\!\!\!\!\!\!\!\!\!\!\!\!\!\!\!\!\!\!\!\!\!\!\!-2\frac{\alpha^2}{N}b(j,m)(r^{22}_{jm+1}-R^{22}_{jm})t. \label{tl4}
\end{eqnarray}

We notice that $\dot{R}^{11}_{jm-1}+\dot{r}^{11}_{jm}+\dot{R}^{22}_{jm}+\dot{r}^{22}_{jm+1}=0.$ This is  a consequence of the  choice of the projection superoperator (Eq.~\eqref{Def-N-J}) as a projection on a conserved quantity of the total system, i.e., the total angular momentum. 

\begin{figure}
\includegraphics[scale=0.6]{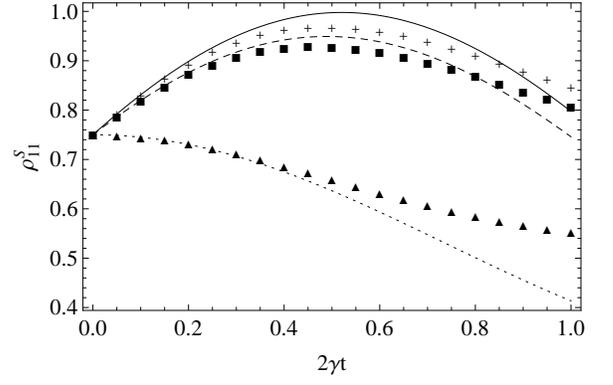}
\caption{Comparison of the dynamics of the population of the upper state $\rho^S_{11}$ occurring in the TCL2 master equation \eqref{tl1}-\eqref{tl4} and the exact solution of the Schr\"{o}dinger equation  \eqref{r11}.   For the initial states $|\Phi^+_{\pi/3,\pi/2}\rangle$  the TCL2 solution is represented by the cross symbol and the exact solution  by the  solid curve. For the initial state $|\Phi^+_{\pi/3,\pi/3}\rangle$  the TCL2 solution is represented by the square symbol and the exact solution  by the  dashed curve. For the initial state $|\Phi^+_{\pi/3,\pi}\rangle$  the TCL2 solution is represented by the triangle symbol and the exact solution  by the  dotted curve.  The coupling strength is $\alpha=\gamma/2$; the  number of spins in the bath is $N=20$.}\label{TCL2_sol}
\end{figure}

The populations can be expressed through the  projections $r$ and $R,$ as
\begin{eqnarray}
\rho^S_{11}=\sum\limits_{j}\sum\limits_{m=-j}^{j}(R^{11}_{jm}+r^{11}_{jm}),\\
\rho^S_{22}=\sum\limits_{j}\sum\limits_{m=-j}^{j}(R^{22}_{jm}+r^{22}_{jm}).
\end{eqnarray} 
The system \eqref{tl1} can be easily solved numerically. We use the convention $r_{j|j+1|}=R_{j|j+1|}=0.$ 
The TCL2 dynamics of the probability to find the   spin in the excited state following from  Eqs. \eqref{tl1}-\eqref{tl4} is shown in Fig. \ref{TCL2_sol}.

\begin{figure}
\includegraphics[scale=0.6]{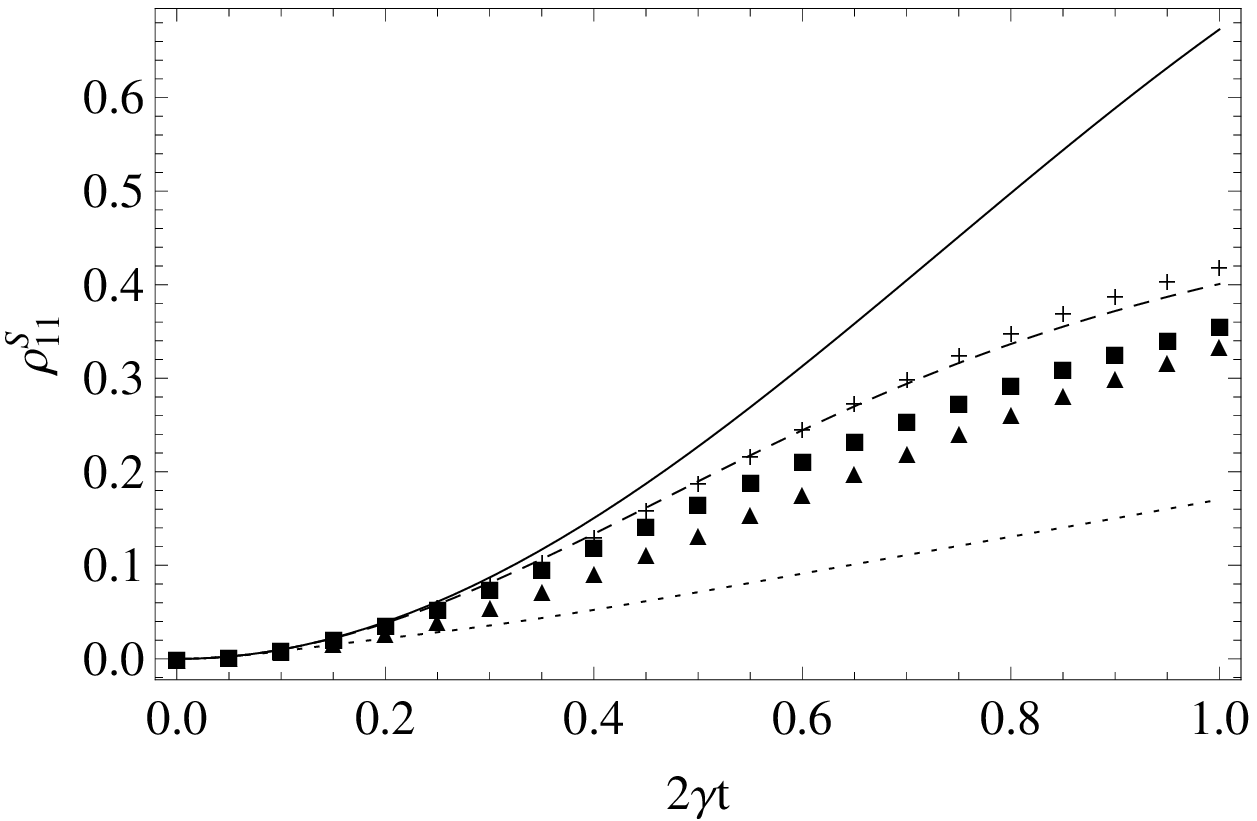}
\caption{Comparison of the dynamics of the population of the upper state $\rho^S_{11}$ occurring in the TCL2 master equation \eqref{tl1}-\eqref{tl4} and the exact solution of the Schr\"{o}dinger equation  \eqref{r11}.  For $\alpha=\gamma/2$  the TCL2 solution is  represented by the cross symbol and the exact solution  by the  solid curve. For $\alpha=2\gamma$ the TCL2 solution is represented by the square symbol and the exact solution by  the dashed curve. For $\alpha=10\gamma$ the TCL2 solution is represented by the triangle symbol and the exact solution by  the dotted curve. The initial state is the Bell state $|\Phi^+_{\pi,0}\rangle$; a number of spins in the bath $N=20$.}\label{TCL2_sol1}
\end{figure}

\begin{figure}
\includegraphics[scale=0.6]{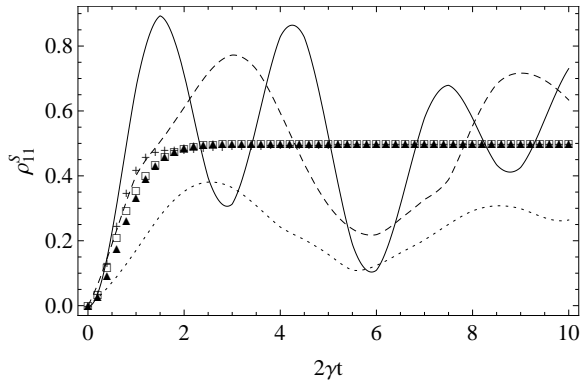}
\caption{Comparison  of the long-time dynamics of the population of the upper state $\rho^S_{11}$ occurring in the TCL2 master equation \eqref{tl1}-\eqref{tl4} and the exact solution of the Schr\"{o}dinger equation  \eqref{r11}.   For $\alpha=\gamma/2$  the TCL2 solution is represented by the cross symbol and the exact solution  by the  solid curve. For $\alpha=2\gamma$ the TCL2 solution is represented by the square symbol and the exact solution by  the dashed curve. For $\alpha=10\gamma$ the TCL2 solution is represented by the triangle symbol and the exact solution by  the dotted curve. The initial state is the Bell state $|\Phi^+_{\pi,0}\rangle$; a number of spins in the bath $N=20$.}\label{TCL2_sol2}
\end{figure}

\section{RESULTS AND DISCUSSION}

The exact dynamics of the   spin is analysed in Figures 2 to 6. Figs. 2 and 3 show the population of the upper state of the  spin in dependence of the strength of the interaction between the bath spins  and the intermediate spin. Figure 2 shows the dynamics for limiting cases. When the strength of the interaction is small, $\alpha\ll\gamma,$ the influence of the bath on the spin dynamics is very weak and the spin evolution is very similar to the case $\alpha=0.$ For the case $\alpha\gg\gamma$ the population oscillates near the initial state with small amplitude. The most interesting case is $\alpha\sim\gamma:$ for this domain we observe a strong dependence on the value of $\alpha$ (see Figs. 2 and 3).

The dynamics of the population of the upper state for different numbers of bath spins is
shown in Figure 4. We can see that the dynamics of the   spin depends very weakly on the number of  spins in the bath. Already for 50 spins a minimal difference from the thermodynamic limit can be observed.

Fig. 5 shows the strong dependence  on  the initial correlation between the reduced spin and the intermediate spin. The initial state of the total system is given by Eq. (4) with $\rho_{SI}(0)$  taken as correlated Bell-like pure states $|\Phi^{+}_{\pi/3,\beta}\rangle$ (Eq.~\eqref{phi}). The different curves in Fig. 4 correspond to  different values of the initial phase parameter $\beta$. Following the explicit expressions for the reduced density matrix in the thermodynamics limit, Eq.~ (27), it is clear that for the model considered here initial system-bath correlations affect the dynamics of the reduced system only if $\Im\rho_{23}(0)\neq 0$. For the initial state analysed in Fig. 5 this corresponds to the case $\beta\neq \pi n \, (n\in \mathbb{Z})$. 

From Fig. 5 one can clearly see that the initial correlations give a non-negligible contribution to the dynamics of the reduced system for all times. The dependence of the dynamics of the reduced system on the initial system bath correlations in the thermodynamic limit is analysed  in Fig. 6. The initial state of the total system is given by Eq. (4) with $\rho_{SI}(0)$ chosen to be correlated Bell-like pure state $|\Phi^{-}_{\pi/3,\beta}\rangle$ (Eq.~\eqref{phi}). Similar to the case presented in Fig. 5, one can see that initial system-bath correlations play an important role in the dynamics of the reduced system. 

Using the explicit formula for the density matrix for the reduced system in the thermodynamic limit, Eq. (27), one can explicitly  analyse limiting cases for the ratio of coupling strengths $\alpha/\gamma$. From Eq. ~(27) one can see that an influence of the bath on the dynamics of the spin is described by the functions $f(t)$  and $g(t)$. In the case $\alpha\ll\gamma$  the expansion of the functions  $f(t)$  and $g(t)$ reads,
$$f(t)\approx\sin^2\frac{\gamma t}{4}+\frac{\alpha^2}{\gamma^2} \left( -2 \sin^2\frac{\gamma t}{4}+\frac{\gamma t}{4} \sin\frac{\gamma t}{2}\right) +O\left( \frac{\alpha^4}{\gamma^4}\right) $$ 
and 
\begin{eqnarray}
\!\!\!\!\!\! g(t)&\!\!\!\!\!\!\!\!\!\!\!\!\!\!\!\!\!\!\!\!\!\!\!\!\!\!\!\!\!\!\!\!\!\!\!\!\!\!\!\!\!\!\!\!\!\!\!\!\!\!\!\!\!\!\!\!\!\!\!\!\approx\frac{1}{2} \exp\left( -\frac{\alpha^2t^2}{8}\right) \left(  \sin\frac{\gamma t}{2}+\right. \nonumber \\
&\left. \frac{\alpha^2}{4\gamma^2}\left( \gamma t \cos\frac{\gamma t}{2}+\left( \frac{\gamma^2 t^2}{4}-2\right) \sin\frac{\gamma t}{2}\right) +O\left( \frac{\alpha^4}{\gamma^4}\right)\right).& \nonumber 
\end{eqnarray}
From the expansion for the functions $f(t)$  and $g(t)$ one can see that in the case $\alpha\ll\gamma$ the dynamics is dominated by the unitary oscillations with the frequency $\gamma/2$ which corresponds to unitary evolution of the spin and intermediate spin.

In the other limiting case, $\alpha\gg\gamma$,  one obtains that $f(t)=g(t)=0$, which explicitly shows the limitation of the equation for the thermodynamic limit  given by the Eq.~\eqref{lim}. As one can see from Fig. 2 in the case $\alpha\gg\gamma$ (Fig. 2 dotted curve) the system is very weakly coupled to an environment and in the limit $\alpha/\gamma\rightarrow \infty$ remains in its initial state.

The limitation of Eq. (26) follows from the asymptotic character of  this equation. This equation gives the correct thermodynamic limit only if the constant of the system bath interaction is small with respect to all the other characteristic parameters of the total system (like in the case $\alpha\ll\gamma$).

Figures 7-9 show the dynamics of the population of the upper state. In order to benchmark the approximation technique the  dynamics of this observable  is derived from the solution of the TCL2 master equation with correlated projection operator  (Eq. \eqref{1} and  Eq. \eqref{2}) and the exact solution of the Schr\"{o}dinger equation (19). The comparison is performed for different initial conditions, parameters $\alpha,\, \gamma$ and different time frames. From  Fig. 7 one can see that  the TCL2 approximation technique  gives good results for sufficiently large time. A deviation  from the exact solution is not exceeded by 5\% for time scale of $~1/\gamma$. In Fig. 8  the dependence on the coupling strength $\gamma/\alpha$ is analysed.  One can see that the correlation projection operator technique gives good agreement for $\gamma/\alpha\gtrsim 1$. The last inequality can be understood if take into account that the equations are perturbative both in alpha and in gamma. The relevant order of the expansion is the maximal value of   two constants $\alpha$ or $\gamma$. However, the evolution of the relevant system defines $\gamma$. For this reason, the most accurate approximation of the dynamics is obtained for  $\alpha\lesssim\gamma$.

 Fig 9. addresses the long time behaviour of the reduced system dynamics. It is clear that for all ranges of the parameters  $\gamma/\alpha$ considered here  the approximation technique shows convergence to a false equilibrium value. However, this behaviour has no correlation with the dynamics described by the exact  solution. From Fig. 9 one can also see that increasing the ratio $\gamma/\alpha$, the discrepancy between approximation technique  and exact solution is growing. A more adequate description of the long time dynamics and parameter ratios $\gamma/\alpha > 1$  might require higher orders in the TCL expansion.

The fact that the TCL2 master equation with correlated  projection operator gives the satisfactory results for time-scales of the order $1/\gamma$ indicates that this approximation technique fits the description of spin-bath systems much better than the traditional form of the projection operator  in the form \eqref{STANDARD-PROJECTION}. As it was indicated in Ref. \cite{oldTCL} a TCL2 master equation for spin systems gives adequate  results only for time scales of the order of $0.05/\gamma$.

For the situation considered in this paper if one chooses a projection operator in the form $\mathcal{P} \rho = (\mathrm{tr}_{IB}\rho) \otimes I_{N+1}/2^{N+1}$, where $N$ is number of bath spins. The corresponding master equation reads
$$\dot{\rho}=-\gamma^2 t/4\left(\sigma^-\sigma^+\rho+\rho\sigma^-\sigma^+-2\sigma^-\rho\sigma^++h.c.\right). $$
It is clear that the above equation does not contain any information about system-bath correlation and cannot give an adequate description of the reduced system dynamics.

In conclusion, we have found an exact solution for a simple spin system coupled to a spin bath through an intermediate spin. We have studied the dynamics of the system and have shown that the initial correlations between the   spin and the intermediate spin have a strong influence on the dynamics of the   spin. On the other hand, the dynamics of the   spin are weakly dependent on the number of bath spins. In addition to the exact solution, an approximate TCL2 master equation was derived with the help of the correlation projection  operator technique. The derived equation explicitly takes into account initial correlations between the   spin and the intermediate spin. The solution of the approximate master equation was compared with the exact solution. It was shown that the approximate technique gives good results for short time dynamics.

\begin{acknowledgments}
This work is based upon research supported by the South African
Research Chair Initiative of the Department of Science and
Technology and National Research Foundation.
\end{acknowledgments}

\end{document}